\documentstyle[preprint,aps,epsfig]{revtex}
\preprint{USM-TH-139}
\begin{document}
\title{Coulomb's law modification in nonlinear and in noncommutative
electrodynamics}
\author{ Patricio Gaete \thanks{E-mail: patricio.gaete@fis.utfsm.cl}
and Iv\'an Schmidt \thanks{E-mail: ivan.schmidt@fis.utfsm.cl}}
\address{Departamento de F\'{\i}sica, Universidad T\'ecnica F.
Santa Mar\'{\i}a, Valpara\'{\i}so, Chile} \maketitle

\begin{abstract}
We study the lowest-order modifications of the static potential
for Born-Infeld electrodynamics and for the $\theta$-expanded
version of the noncommutative $U(1)$ gauge theory, within the
framework of the gauge-invariant but path-dependent variables
formalism. The calculation shows a long-range correction
($1/r^5$-type) to the Coulomb potential in Born-Infeld
electrodynamics. However, the Coulomb nature of the potential ( to
order $e^2$ ) is preserved in noncommutative electrodynamics.
\end{abstract}
\smallskip

PACS number(s): 11.10.Ef, 11.10.Nx

\section{INTRODUCTION}

Renewed interest in non-linear electrodynamics (Born-Infeld
theory) has originated in recent string theory investigations.
This is primarily because the low energy dynamics of D-branes have
been  described by a nonlinear Born-Infeld type
action\cite{Tseytlin,Gibbons}. It is worth recalling at this stage
that Born and Infeld \cite{Born} suggested to modify Maxwell's
electromagnetism so as to get rid of the divergencies of the
theory such as the infinite self-energy of a point charge. The
resulting theory is a nonlinear gauge theory endowed with
interesting features, like finite electron self-energy and a
regular point charge electric field at the origin. However, due to
the nonlinearity, the corresponding field equations are very
difficult to solve. In addition to the string interest, the
Born-Infeld theory has also attracted considerable attention from
different viewpoints. For example, in connection to duality
symmetry\cite{Rivelles,Khoudeir}, also in magnetic monopoles
studies\cite{Kim}, in generation of multipole moments for charged
particles\cite{Chruscinski}, and possible experimental
determination of the parameter that measures the nonlinearity of
the theory\cite{Denisov}. The advent of noncommutative field
theories also called attention to nonlinear theories. In fact,
recently there have been indications that the $\theta$-expanded
version of the noncommutative U(1) gauge theory is equivalent to
the expansion of the Born-Infeld action up to order
$F^3$\cite{Gomis}, which is something that we intend to check in
the present work, for the specific case of the static potential
between charges.

On the other hand, one may improve our understanding of gauge
theories through a proper study of the concepts of screening and
confinement. In this respect the interaction energy of an
infinitely heavy quark-antiquark pair is a key tool which plays an
important role in the understanding of quark confinement.
Moreover, the static potential is an essential concept both in
electrodynamics and in gravitation. The importance of the static
potential is further shown, for example, in the description of
non-relativistic bound systems like quarkonia, as well as in the
definition of the lattice coupling. In actual calculations it is
obtained most directly when a correct separation between
gauge-invariant and gauge dependent degrees of freedom is made.
Previously, we proposed a general framework for studying the
confining and screening nature of the potential in gauge theories
in terms of gauge-invariant but path-dependent field variables
\cite{Gaete}. An important feature of this methodology is that it
provides a physically-based alternative to the usual Wilson loop
approach. In this paper, we will examine another aspect of
nonlinear theories, namely, the lowest-order modification of the
static potential due to the presence of Born-Infeld type terms,
and in noncommutative electrodynamics. We address this issue along
the lines of reference \cite{Gaete}.

In Sec. II we will calculate the lowest-order correction to the
Coulomb energy of a fermion-antifermion system, for both
Born-Infeld electrodynamics and for the $\theta$-expanded version
of noncommutative U(1) gauge theory. Our calculations show that
the static interaction of fermions in Abelian gauge theories is
determined by the geometrical condition of gauge invariance.

\section{Interaction Energy}
\subsection{Two-dimensional Born-Infeld Electrodynamics}

As already stated, our principal purpose is to calculate
explicitly the interaction energy between static pointlike sources
for Born-Infeld electrodynamics. To this end  we will calculate
the expectation value of the energy operator $ H$ in the physical
state $ |\Phi\rangle$, which we will denote by ${ \langle H
\rangle_ \Phi}$. However, before going into the four-dimensional
Born-Infeld electrodynamics, we would like to first consider the
two-dimensional case. This would not only provide the theoretical
setup for our subsequent work, but also fix the notation. The
starting point is the two-dimensional space-time Lagrangian:
\begin{equation}
{\cal L} = \beta ^2 \left\{ {1 - \sqrt {1 + \frac{1}{{2\beta ^2
}}F_{\mu \nu } F^{\mu \nu } } } \right\} - A_0 J^0,\label{bid1}
\end{equation}
where $J^0$ is the external current. The parameter $ \beta $
measures the nonlinearity of the theory and in the limit $ \beta
\to \infty $ the Lagrangian (\ref{bid1}) reduces to the Maxwell
theory. In order to handle the square root in the Lagrangian
(\ref{bid1}) we introduce an auxiliary field $v$, such that its
equation of motion gives back the original theory\cite{Tseytlin}.
This allows us to write the Lagrangian as
\begin{equation}
{\cal L} = \beta ^2 \left\{ {1 - \frac{v}{2}\left( {1 +
\frac{1}{{2\beta ^2 }}F_{\mu \nu } F^{\mu \nu } } \right) -
\frac{1}{{2v}}} \right\} - A_0 J^0. \label{bid2}
\end{equation}
Once this is done, the canonical quantization of this theory from
the Hamiltonian analysis point of view is straightforward and
follows closely that of references \cite{Rivelles,Khoudeir}. The
canonical momenta read $\Pi ^\mu   =  - vF^{0\mu }$, and one
immediately identifies the two primary constraints $ \Pi ^0  = 0 $
and $ p \equiv \frac{{\partial {\cal L}}}{{\partial v}} = 0$. The
canonical Hamiltonian is thus
\begin{equation}
H_C  = \int {dx\left\{ { - \beta ^2  + \Pi _1 \partial ^1 A^0  +
\frac{1}{{2v}}\left( {\beta ^2  - \Pi _1 \Pi ^1 } \right) +
\frac{{\beta ^2 }}{2}v + A_0 J^0 } \right\}}. \label{bid3}
\end{equation}
The consistency condition ${\dot \Pi _0}=0$ leads to the secondary
constraint $\Gamma _1 \left( x \right) \equiv \partial _1 \Pi ^1 -
J^0=0$. The consistency condition for the $p$ constraint yields no
further constraints and just determines the field $v$,
\begin{equation}
v = \sqrt {1 - \frac{1}{{\beta ^2 }}\Pi _1 \Pi ^1 }, \label{bid4}
\end{equation}
which will be used to eliminate $v$. The extended Hamiltonian that
generates translations in time then reads $H = H_C  + \int d x
\left( {c_0 (x)\Pi_0 (x) + c_1 (x)\Gamma _1 (x)} \right)$, where
$c_0(x)$ and $c_1(x)$ are the Lagrange multipliers. Since $ \Pi_0
= 0$ for all time and $ \dot{A}_0 \left( x \right) = \left[ {A_0
\left( x \right),H} \right] = c_0 \left( x \right)$, which is
completely arbitrary, we discard $ A_0 \left( x \right)$ and $ \Pi
_0 \left( x \right)$ because they add nothing to the description
of the system. Then, the Hamiltonian takes the form
\begin{equation}
H = \int {dx\left\{ {\beta ^2 \left( {\sqrt {1 - \frac{1}{{\beta
^2 }}\Pi _1 \Pi ^1 }  - 1} \right) - c^ \prime  \left( x
\right)\left( {\partial _1 \Pi ^1  - J^0 } \right)} \right\}},
\label{bid5}
\end{equation}
where $c^ \prime  \left( x \right) = c_1 \left( x \right) - A_0
\left( x \right)$.

The quantization of the theory requires the removal of
non-physical variables, which is done by imposing a gauge
condition such that the full set of constraints becomes second
class. A convenient choice is found to be \cite{Gaete}
\begin{equation}
\Gamma _2 \left( x \right) \equiv \int\limits_{C_{\xi x} } {dz^\nu
} A_\nu \left( z \right) \equiv \int\limits_0^1 {d\lambda x^1 }
A_1 \left( {\lambda x} \right) = 0, \label{bid6}
\end{equation}
where  $\lambda$ $(0\leq \lambda\leq1)$ is the parameter
describing the spacelike straight path $ x^1  = \xi ^1  + \lambda
\left( {x - \xi } \right)^1 $, and $ \xi $ is a fixed point
(reference point). There is no essential loss of generality if we
restrict our considerations to $ \xi ^1=0 $. With this choice the
nontrivial Dirac bracket is given by
\begin{equation}
\left\{ {A_1 \left( x \right),\Pi ^1 \left( y \right)} \right\}^ *
= \delta ^{\left( 1 \right)} \left( {x - y} \right) -
\partial _1^x \int\limits_0^1 {d\lambda x^1 } \delta ^{\left( 1
\right)} \left( {\lambda x - y} \right). \label{bid7}
\end{equation}

Having outlined the necessary aspects of quantization, we now turn
to the problem of obtaining the interaction energy between
pointlike sources in Born-Infeld theory, where a fermion is
localized at the origin $ {\bf 0}$ and an antifermion at $ {\bf
y}$. As we have already indicated, we will calculate the
expectation value of the energy operator $ H$ in the physical
state $ |\Phi\rangle$. From (\ref{bid5}) we then get for the
expectation value
\begin{equation}
\left\langle H \right\rangle _\Phi   = \left\langle \Phi
\right|\int {dx\left\{ {\beta ^2 \left( {\sqrt {1 -
\frac{1}{{\beta ^2 }}\Pi _1 \Pi ^1 }  - 1} \right)} \right\}}
\left| \Phi  \right\rangle. \label{bid8}
\end{equation}
As remarked by Dirac\cite{Dirac}, the physical state can be
written as
\begin{equation}
\left| \Phi  \right\rangle  \equiv \left| {\overline \Psi  \left(
\bf y \right)\Psi \left( \bf 0 \right)} \right\rangle  = \overline
\psi \left( \bf y \right)\exp \left( {ie\int\limits_{\bf 0}^{\bf
y} {dz^i } A_i \left( z \right)} \right)\psi \left(\bf 0
\right)\left| 0 \right\rangle, \label{bid9}
\end{equation}
where $\left| 0 \right\rangle$ is the physical vacuum. As before,
the line integral appearing in the above expression is along a
spacelike path starting at $\bf 0$ and ending at $\bf y$, on a
fixed time slice. It is worth noting here that the strings between
fermions have been introduced in order to have a gauge-invariant
function $ \left| \Phi \right\rangle $. In other terms, each of
these states represents a fermion-antifermion pair surrounded by a
cloud of gauge fields sufficient to maintain gauge invariance.

Since we are interested in estimating the lowest-order correction
to the Coulomb energy, we will retain only the leading quadratic
term in the expression (\ref{bid8}). Thus the expectation value
simplifies to
\begin{equation}
\left\langle H \right\rangle _\Phi   = \left\langle \Phi
\right|\int {dx} \left\{ {\frac{1}{2}\left( {\Pi _1 } \right)^2  -
\frac{1}{{8\beta ^2 }}\left( {\Pi _1 } \right)^4 } \right\}\left|
\Phi  \right\rangle. \label{bid10}
\end{equation}
It is easy to see that the first term inside the curly bracket
comes from the usual Maxwell theory while the second one is a
correction which comes from the Born-Infeld modification. From our
above Hamiltonian analysis we observe that
\begin{equation}
\Pi _1 \left( x \right)\left| {\overline \Psi  \left( y
\right)\Psi \left( 0 \right)} \right\rangle  = \overline \Psi
\left( y \right)\Psi \left( 0 \right)\Pi _1 \left( x \right)\left|
0 \right\rangle  - e\int_0^y {dz_1 } \delta ^{\left( 1 \right)}
\left( {z_1  - x} \right)\left| \Phi  \right\rangle. \label{bid11}
\end{equation}
Substituting this back into (\ref{bid10}), we obtain
\begin{equation}
\left\langle H \right\rangle _\Phi   = \left\langle H
\right\rangle _0  + \frac{{e^2 }}{2}\int {dx} \left( {\int_0^y
{dz_1 \delta ^{\left( 1 \right)} \left( {z_1  - x} \right)} }
\right)^2  - \frac{{e^4 }}{{8\beta ^2 }}\int {dx} \left( {\int_0^y
{dz_1 \delta ^{\left( 1 \right)} \left( {z_1  - x} \right)} }
\right)^4, \label{bid12}
\end{equation}
where $\left\langle H \right\rangle _0  = \left\langle 0
\right|H\left| 0 \right\rangle$. We further note that
\begin{equation}
\frac{{e^2 }}{2}\int {dx} \left( {\int_0^y {dz\delta ^1 \left(
{z_1  - x} \right)} } \right)^2  = \frac{{e^2 }}{2}L ,
\label{pato}
\end{equation}
with $|y|\equiv L$. By employing Eq. (\ref{pato}) we can reduce
Eq. (\ref{bid12}) to
\begin{equation}
V = \frac{{e^2 }}{2}\left( {1 - \frac{{e^2 }}{{4\beta ^2 }}}
\right)L. \label{bid13}
\end{equation}
Hence we see that the static interaction between fermions arises
only because of the requirement that the $\left| {\overline \Psi
\Psi } \right\rangle$ states be gauge invariant. The above result
reveals that the effect of adding the Born-Infeld term is to
decrease the energy. Nevertheless, the confining nature of the
potential is preserved.

Eq. (\ref{bid13}) exhibits the same formal structure as the one
obtained for the massive Schwinger model. In fact, the same
calculation for the massive Schwinger model \cite{Gaete} gives
\begin{equation}
V = \frac{{q^2 }}{2}\left( {1 + \frac{{e^2 }}{{4\pi ^2 m\Sigma }}}
\right)^{ - 1} L , \label{msm}
\end{equation}
where $\Sigma  = \left( {\frac{e}{{2\pi ^{\frac{3}{2}} }}}
\right)\exp \left( {\gamma _E } \right)$ with ${\gamma _E }$ the
Euler-Mascheroni constant, $m$ and $e$ are the mass and charge of
the dynamical fermions. Here $q$ refers to the probe charges.
Considering the limit $m \gg e$ and $q \equiv e$, we get
\begin{equation}
V = \frac{{e^2 }}{2}\left( {1 - \frac{{e^2 }}{{4\pi ^2 m\Sigma }}}
\right)L . \label{msm2}
\end{equation}
Therefore, for this special case the massive Schwinger model
simulates the features of the Born-Infeld theory. This analysis
suggests the interesting possibility of identifying the parameter
$\beta$ with the mass of the dynamical fermions. This, however, is
a separate question and which we do not intend to address here.

Before concluding this subsection we discuss an alternative
derivation of the result (\ref{bid13}), which highlights certain
distinctive features of our methodology. We start by considering
\begin{equation}
V \equiv e\left( {{\cal A}_0 \left( \bf 0 \right) - {\cal A}_0
\left( \bf y \right)} \right), \label{bid14}
\end{equation}
where the physical scalar potential is given by
\begin{equation}
{\cal A}_0 \left( {x^0 ,x^1 } \right) = \int_0^1 {d\lambda } x^1
E_1 \left( {\lambda x^1 } \right). \label{bid15}
\end{equation}
This follows from the vector gauge-invariant field
expression\cite{GaeteB}
\begin{equation}
{\cal A}_\mu  \left( x \right) \equiv A_\mu  \left( x \right) +
\partial _\mu  \left( { - \int_\xi ^x {dz^\mu  A_\mu  \left( z
\right)} } \right), \label{bid16}
\end{equation}
where, as in Eq.(\ref{bid6}), the line integral is along a
spacelike path from the point $\xi$ to $x$, on a fixed time slice.
The gauge-invariant variables (\ref{bid16}) commute with the sole
first constraint (Gauss' law), confirming that these fields are
physical variables \cite{Dirac}. Note that Gauss' law for the
present theory reads
\begin{equation}
\partial _1 \frac{{E^1 }}{{\sqrt {1 - \frac{1}{{\beta ^2 }}
\left( {E^1 } \right)^2 } }} = J^0, \label{bid17}
\end{equation}
where $E^1$ is the one-dimensional electric field. For $J^0 \left(
{t,x} \right) = e\delta ^{\left( 1 \right)} \left( x \right)$, the
electric field is given by
\begin{equation}
E^1  = \frac{e}{2}\frac{1}{{\sqrt {1 + \frac{{e^2 }}{{4\beta ^2
}}} }}\hat x^1, \label{bid18}
\end{equation}
here $\hat x^1$ is an unit vector ($\hat x^1 = \frac{{x^1
}}{{|x^1|}}$). Finally, making use of (\ref{bid18}) and
(\ref{bid15}) in (\ref{bid14}), we find
\begin{equation}
V = \frac{{e^2 }}{2}\left( {1 - \frac{{e^2 }}{{4\beta ^2 }}}
\right)L, \label{bid19}
\end{equation}
where $|y|\equiv L$.

\subsection{Four-dimensional Born-Infeld Electrodynamics}

We now turn our attention to the calculation of the interaction
energy between static pointlike sources for the four-dimensional
Born-Infeld electrodynamics. In such a case the Lagrangian reads
\begin{equation}
{\cal L} = \beta ^2 \left\{ {1 - \sqrt {1 + \frac{1}{{2\beta ^2
}}F^{\mu \nu } F_{\mu \nu }  - \frac{1}{{16\beta ^4 }}\left(
{{\cal F}^{\mu \nu } F_{\mu \nu } } \right)^2 } } \right\} - A_0
J^0.\label{borin1}
\end{equation}
where $ \frac{1}{4}F_{\mu \nu } F^{\mu \nu }  =  -
\frac{1}{2}\left( {{\bf E}^2  - {\bf B}^2 } \right)$ , $ F_{\mu
\nu } {\cal F}^{\mu \nu }  = 4\left( {{\bf E} \cdot {\bf B}}
\right)$ and $J^0$ is the external current.

Before we proceed to work out explicitly the energy, we shall
begin by summarizing the Hamiltonian analysis of the theory
(\ref{borin1}). Once again, we will introduce an auxiliary field
$v$ to handle the square root in the Lagrangian (\ref{borin1}).
Expressed in terms of this field, the Lagrangian (\ref{borin1})
takes the form
\begin{equation}
{\cal L} = \beta ^2 \left\{ {1 - \frac{v}{2}\left( {1 +
\frac{1}{{2\beta ^2 }}F_{\mu \nu } F^{\mu \nu }  -
\frac{1}{{16\beta ^4 }}\left( {{\cal F}^{\mu \nu } F_{\mu \nu } }
\right)^2 } \right) - \frac{1}{{2v}}} \right\} - A^0
J_0.\label{borin2}
\end{equation}
With this in hand, the canonical momenta are $ \Pi ^\mu   =  -
v\left( {F^{0\mu }  - \frac{1}{{4\beta ^2 }}F_{\alpha \beta }
{\cal F}^{\alpha \beta } {\cal F}^{0\mu } } \right)$, and one
immediately identifies the two primary constraints $ \Pi ^0  = 0 $
and $ p \equiv \frac{{\partial {\cal L}}}{{\partial v}} = 0$. The
canonical Hamiltonian of the model can be worked out as usual and
is given by the expression
\begin{equation}
H_C  = \int {d^3 x} \left\{ { - \beta ^2  +  \Pi _i \partial ^i
A^0 + \frac{1}{{2v}}\left( {{\bf \Pi} ^2  + \beta ^2 } \right) +
\frac{v}{2}\left( {{\bf B}^2  + \beta ^2 } \right) -
\frac{1}{{2v\beta ^2 }}\frac{{\left( {{\bf \Pi}  \cdot {\bf B}}
\right)^2 }}{{\left( {1 + \frac{{{\bf B}^2 }}{{\beta ^2 }}}
\right)}} + A_0 J^0 } \right\} . \label{borin3}
\end{equation}
Requiring the primary constraint $\Pi _0 $ to be preserved in time
yields the secondary constraint (Gauss' law) $\Gamma _1 \left( x
\right) \equiv\partial _i \Pi ^i - J^0=0$. Similarly for the
constraint $p$, we get the auxiliary field $v$ as
\begin{equation}
v = \frac{1}{{\beta ^2 \left( {1 + \frac{1}{{\beta ^2 }}{\bf B}^2
} \right)}}\sqrt {\beta ^2 \left( {{\bf \Pi} ^2  + \beta ^2 }
\right)\left( {1 + \frac{1}{{\beta ^2 }}{\bf B}^2 } \right) -
\left( {{\bf \Pi}  \cdot {\bf B}} \right)^2 } . \label{borin4}
\end{equation}
The extended Hamiltonian that generates translations in time then
reads $H = H_C  + \int {d^3 } x\left( {c_0 (x)\Pi_0 (x) + c_1
(x)\Gamma _1 (x)} \right)$, where $c_0(x)$ and $c_1(x)$ are
Lagrange multipliers. As before, neither $ A_0 \left( x \right)$
nor $ \Pi _0 \left( x \right)$ are of interest in describing the
system and may be discarded from the theory. Thus we are left with
the following expression for the Hamiltonian
\begin{equation}
H = \int {d^3 x} \left\{ {\sqrt {\beta ^2 \left( {{\bf \Pi} ^2  +
\beta ^2 } \right)\left( {1 + \frac{{{\bf B}^2 }}{{\beta ^2 }}}
\right) - \left( {{\bf \Pi}  \cdot {\bf B}} \right)^2 }  - \beta
^2 + c^{\prime} (x)\left( {\partial _i \Pi ^i  - J^0 } \right)}
\right\}, \label{borin5}
\end{equation}
where $ c^{\prime }(x)=c_{1}(x)-A_{0}(x)$.

Since our main motivation is to compute the static potential for
the Born-Infeld theory, we will adopt the same gauge-fixing
condition that was used in the last subsection, that is,
\begin{equation}
\Gamma _2 \left( x \right) \equiv \int\limits_{C_{\xi x} } {dz^\nu
} A_\nu \left( z \right) \equiv \int\limits_0^1 {d\lambda x^i }
A_i \left( {\lambda x} \right) = 0. \label{borin6}
\end{equation}
Here again $\lambda$ $(0\leq \lambda\leq1)$ is the parameter
describing the spacelike straight path $ x^i  = \xi ^i  + \lambda
\left( {x - \xi } \right)^i $ with $i=1,2,3$,  and $ \xi $ is a
fixed (reference) point. There is no essential loss of generality
if we restrict our considerations to $ \xi ^i=0 $. In this way,
the fundamental Dirac bracket can be rewritten as
\begin{equation}
\left\{ {A_i \left( x \right),\Pi ^j \left( y \right)} \right\}^ *
= \delta _i^j \delta ^{\left( 3 \right)} \left( {x - y} \right) -
\partial _i^x \int\limits_0^1 {d\lambda x^i } \delta ^{\left( 3
\right)} \left( {\lambda x - y} \right). \label{borin7}
\end{equation}

We are now in a position to compute the potential energy for
static charges in this theory. To do this, we will use the
gauge-invariant scalar potential which is now given by
\begin{equation}
{\cal A}_0 (t,{\bf r}) = \int_0^1 {d\lambda } r^i E_i (t,\lambda
{\bf r}). \label{borin8}
\end{equation}
It follows from the above discussion that Gauss' law takes the
form
\begin{equation}
\partial _i \frac{{E_i }}{{\sqrt {1 - \frac{{´{\bf E}^2 }}{{\beta ^2
}}} }} = J^0. \label{borin9}
\end{equation}
For $J^0 \left( {t,{\bf r}} \right) = e\delta ^{\left( 3 \right)}
\left( {\bf r} \right)$, the electric field follows as
\begin{equation}
E^i \left( {\bf r} \right) = \frac{e}{{4\pi }}\frac{1}{{\sqrt
{|{\bf r}|^4 + \rho _0^2 } }}\hat r^i, \label{borin10}
\end{equation}
where $ \rho _0  \equiv \frac{e}{{4\pi \beta }} $ and $ \hat r^i =
\frac{{{\bf r}^i }}{{|{\bf r}|}} $. From this expression it should
be clear that the electric field of a pointlike charge is regular
at the origin, in contrast to the usual Maxwell theory. As a
consequence, equation (\ref{borin8}) becomes
\begin{equation}
{\cal A}_0 \left( {t,{\bf r}} \right) =  - \frac{e}{{4\pi
}}\int_0^1 {d\lambda } \frac{r}{{\sqrt {\left( {\lambda r}
\right)^4  + \rho _0^2 } }}.  \label{borin11}
\end{equation}
Again, as in the two-dimensional case, it is sufficient to retain
the leading quadratic term in Eq. (\ref{borin11}). Thus we obtain
\begin{equation}
{\cal A}_0 \left( {t,{\bf r}} \right) =  - \frac{e}{{4\pi
r}}\int_0^1 {d\lambda } \left\{ {\frac{1}{{\lambda ^2 }} +
\frac{1}{2}\frac{{a^4 }}{{\lambda ^6 }}} \right\}, \label{borin12}
\end{equation}
where $ a^4  \equiv \frac{{\rho _0^2 }}{{r^4 }} = \frac{{e^2
}}{{16\beta ^2 \pi ^2 r^4 }} $.

In terms of $ {\cal A}_0 \left( {t,{\bf r}} \right) $, the
potential for a pair of static pointlike opposite charges located
at $\bf 0$ and $\bf L$, that is, $ J^0 \left( {t,{\bf r}} \right)
= e\left\{ {\delta ^{\left( 3 \right)} \left( {\bf r} \right) -
\delta ^{\left( 3 \right)} \left( {{\bf r} - {\bf L}} \right)}
\right\}$, is given by
\begin{equation}
V \equiv e \left( {{\cal A}_0 \left( {\bf 0} \right) - {\cal A}_0
\left( {\bf L} \right)} \right) =  - \frac{{e^2 }}{{4\pi
}}\frac{1}{L}\left( {1 + \frac{{e^2 }}{{160\pi ^2 \beta ^2
}}\frac{1}{{L^4 }}} \right), \label{borin13}
\end{equation}
with $ L = |{\bf L}|$.
\begin{figure}
\begin{center}
\epsfig{figure=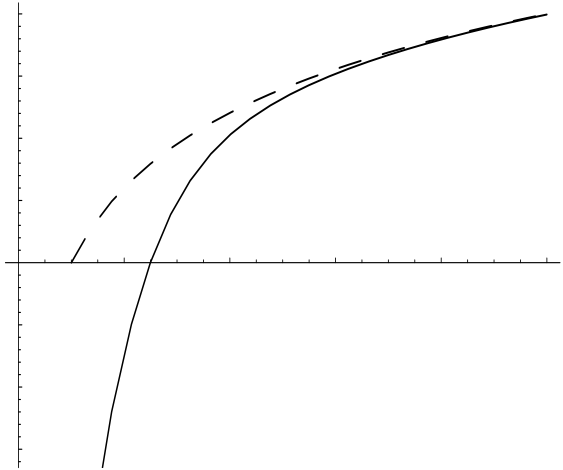 , height=8cm, width=9cm} 
\caption {Shape of $Log$ $V(L)$ (in units of
$\alpha\equiv \frac{e^2}{{4\pi }}$), as a function of the distance
$L$. The dashed line represents the Coulomb potential (in units of
$\alpha\equiv \frac{e^2}{{4\pi }}$).}
\end{center}
\end{figure}

This result shows the usual Coulomb potential with a long-range
correction due to the second term of the form
\begin{equation}
| \Delta  V |  =  \frac{{e^4 }}{{640\pi ^3 \beta ^2
}}\frac{1}{{L^5 }} . \label{correc}
\end{equation}

To ${\cal O}\left( {\frac{1}{{\beta ^2 }}} \right)$ Born-Infeld
electrodynamics displays a marked qualitative departure from the
usual Maxwell theory. In Fig.$1$ we show the effect of the
${\beta}$ -correction, for the case $\beta=10$. Note that we have
plotted the logarithm of $V(L)$ as a function of $L$. It is
important to notice that the presence of the second term on the
right-hand side of Eq. (\ref{borin13}), which dominates for small
$L$ values, causes $V$ to decrease. Thus, from a physical point of
view, this discussion allows us to say that the effect of adding
the Born-Infeld term is to generate more stable bound states of
charged particles. Let us also mention here that if we had
considered the value of $\beta$ as predicted in \cite{Denisov},
the correction to the Coulomb potential would have been
negligible. Hence it becomes important to obtain an independent
lower bound for $\beta$ from our long-range correction, which we
hope to report elsewhere.

\subsection{Non-commutative electrodynamics}

We now want to extend what we have done to non-commutative
electrodynamics to leading order in $\theta$. As before, we will
concentrate on the effect of including the $\theta$ term in the
static potential. In such a case the Lagrangian
\cite{Seiberg,Grimstru,Jackiw} reads
\begin{equation}
{\cal L} =  - \frac{1}{4}F_{\mu \nu }^2  + \frac{1}{8}\theta
^{\alpha \beta } F_{\alpha \beta } F_{\mu \nu }^2  -
\frac{1}{2}\theta ^{\alpha \beta } F_{\mu \alpha } F_{\nu \beta }
F^{\mu \nu }  - A_0 J^0 , \label{ncmm1}
\end{equation}
( $\mu,\nu,\alpha,\beta=0,1,2,3$ ). Here, the field-strength
tensor is expressed in terms of potentials in the usual way, $
F_{\mu \nu } = \partial _\mu  A_\nu   - \partial _\nu  A_\mu$, and
$J^0$ is an external current. As is well known $\theta ^{\alpha
\beta }$ is a real constant antisymmetric tensor, and from now on
we take $ \theta ^{0 \alpha}=0 $ and $ \theta ^{ij}= \varepsilon
^{ijk}\theta ^{k}$.

The above Lagrangian will be the starting point of the Dirac
constrained analysis \cite{Kruglov}. The canonical momenta
following from Eq. (\ref{ncmm1}) are $ \Pi ^\mu   = \left( {1 -
\frac{1}{2}\theta ^{\alpha \beta } F_{\alpha \beta } }
\right)F^{\mu 0}  - \theta ^{\mu \beta } F_{\nu \beta } F^{0\nu }
- \theta _{\alpha \beta } F^{0\alpha } F^{\mu \beta }$, which
results in the usual primary constraint $\Pi ^0=0$ and $\Pi ^i  =
\left( {1 - \frac{1}{2}\theta ^{kl} F_{kl} } \right)F^{i0}  -
\theta ^{il} F_{jl} F^{0j}  - \theta _{kl} F^{0k} F^{il}$ (
$i,j,k,l=1,2,3$ ). Defining the electric and magnetic fields by $
E^i  = F^{i0}$ and $B^i  = \frac{1}{2}\varepsilon ^{ijk} F_{jk}$,
respectively, the canonical Hamiltonian assumes the form
\begin{equation}
H_C  = \int {d^3 x}   \left\{ {\frac{1}{2}\left( {{\bf E}^2 + {\bf
B}^2 } \right)\left( {1 + \mbox{\boldmath$\theta$}  \cdot {\bf B}}
\right) - \left( {\mbox{\boldmath$\theta$}  \cdot {\bf E}}
\right)\left( {{\bf E} \cdot {\bf B}} \right) - A_0 \left(
{\partial _i \Pi ^i - J^0 } \right)} \right\} .  \label{ncmm2}
\end{equation}
Time conservation of the primary constraint leads to the secondary
constraint $\Gamma_1(x) \equiv \partial_i\Pi^i - J^0=0$, and the
time stability of the secondary constraint does not induce more
constraints, which are first class. It should be noted that the
constrained structure for the gauge field remains identical to the
Born-Infeld theory. Thus, the quantization can be done in a
similar manner to that in the previous subsection. In view of this
situation, we pass now to the calculation of the interaction
energy.

Following our earlier procedure, we will compute the expectation
value of the noncommutative electrodynamics Hamiltonian in the
physical state $\left| \Phi  \right\rangle$ (Eq. (\ref{bid9})).
That is,
\begin{equation}
\left\langle H \right\rangle _\Phi   = \left\langle \Phi
\right|\int {d^3 x} \left\{ {\frac{1}{2}\left( {{\bf E}^2  + {\bf
B}^2 } \right)\left( {1 + \mbox{\boldmath$\theta$}  \cdot {\bf B}}
\right) - \left( {\mbox{\boldmath$\theta$} \cdot {\bf B}}
\right)\left( {{\bf E} \cdot {\bf B}} \right)} \right\}\left| \Phi
\right\rangle. \label{ncmm3}
\end{equation}
From our above Hamiltonian analysis, Eq. (\ref{ncmm3}) can be
simplified
\begin{equation}
\left\langle H \right\rangle _\Phi   = \left\langle \Phi
\right|\int {d^3 } x\left\{ {\frac{1}{2}{\ {\bf E}}^2 \left( {1 +
\mbox{\boldmath$\theta$} \cdot {\bf B}} \right) - \left(
{\mbox{\boldmath$\theta$} \cdot {\bf B}} \right)\left( {{\bf E}
\cdot {\bf B}} \right)} \right\}\left| \Phi \right\rangle
.\label{ncmm4}
\end{equation}
According to the definition of the canonical momenta $\Pi^i$, we
may also write
\begin{equation}
E^i  = \left( {1 + \mbox{\boldmath$\theta$}  \cdot {\bf B}}
\right)\Pi ^i - \left( {\mathbf{\mbox{\boldmath$\theta$}  \cdot
\Pi}} \right)B^i - \theta ^i \left( \mathbf{\Pi \cdot B} \right) ,
\label{ncmm5}
\end{equation}
to lowest order in $\theta$. Using Eq. (\ref{ncmm5}) we can
rewrite Eq. (\ref{ncmm4}) in the following way
\begin{equation}
\left\langle H \right\rangle _\Phi   = \left\langle \Phi
\right|\int {d^3 } x\left\{ {\frac{1}{2}{\mathbf\Pi} ^2  +
\frac{3}{2}\left( {\mbox{\boldmath$\theta$}  \cdot {\bf B}}
\right){\mathbf\Pi} ^2 - 3\left( {\mathbf{\mbox{\boldmath$\theta$}
\cdot \Pi} } \right)\left({\mathbf {\Pi \cdot B}} \right)}
\right\}\left| \Phi \right\rangle . \label{ncmm6}
\end{equation}
Taking into account the preceding Hamiltonian structure, we first
note that
\begin{equation}
\Pi _i \left( {\bf x} \right) = \overline \psi  \left( {\bf y}
\right)\psi \left( {\bf 0} \right)\Pi _i \left( {\bf x}
\right)\left| 0 \right\rangle  + e\int_{\bf 0}^{\bf y} {dz_i }
\delta ^{\left( 3 \right)} \left( {{\bf x} - {\bf z}}
\right)\left| \Phi \right\rangle . \label{pato2}
\end{equation}
Using this in (\ref{ncmm6}) we then evaluate the expectation value
in the presence of the static charges
\begin{equation}
\left\langle H \right\rangle _\Phi   = \left\langle H
\right\rangle _0  + V_1  + V_2^\theta   + V_3^\theta ,
\label{ener1}
\end{equation}
where $\left\langle H \right\rangle _0  = \left\langle 0
\right|H\left| 0 \right\rangle$. The $V_1, V_2^\theta, V_3^\theta$
terms are given by
\begin{equation}
V_1  = \frac{{e^2 }}{2}\int {d^3 } x\left( {\int_{\bf 0}^{\bf y}
{dz_i \delta ^{\left( 3 \right)} \left( {{\bf z} - {\bf x}}
\right)} } \right)^2 , \label{ener2}
\end{equation}
\begin{equation}
V_1^\theta   = \frac{3}{2}e^2 \int {d^3 x} \left(
{\mbox{\boldmath$\theta$} \cdot {\bf B}\left( {\bf x} \right)}
\right)\left( {\int_{\bf 0}^{\bf y} {dz_i \delta ^{\left( 3
\right)} \left( {{\bf z} - {\bf x}} \right)} } \right)^2 ,
\label{ener3}
\end{equation}
\begin{equation}
V_2^\theta   =  - 3e^2 \int {d^3 } x\left( {\theta _i B_j \left(
{\bf x} \right)} \right)\int_{\bf 0}^{\bf y} {dz_i } \delta
^{\left( 3 \right)} \left( {{\bf z} - {\bf x}} \right)\int_{\bf
0}^{\bf y} {dz_j^ \prime  } \delta ^{\left( 3 \right)} \left(
{{\bf z}^ \prime - {\bf x}} \right) . \label{ener4}
\end{equation}
The integrals over $z_i$ and $z^{\prime}_j$ are zero except on the
contour of integration.

Here we make the following observations. First, we note that the
term (\ref{ener2}) may look peculiar, but it is nothing but the
familiar Coulomb interaction plus a self-energy term. In effect,
as was explained in \cite{GaeteB}, by using spherical coordinates
the integral $\int_{\bf 0}^{\bf y} {dz_i } \delta ^{\left( 3
\right)} \left( {{\bf z} - {\bf x}} \right)$ can also be written
as
\begin{equation}
\int_{\bf 0}^{\bf y} {dz_i } \delta ^3 \left( {{\bf x} - {\bf z}}
\right) = \frac{{{\bf y}_i }}{{|{\bf y}|}}\frac{1}{{|{\bf y} -
{\bf x}|^2 }}\sum\limits_{l,m} {Y_{lm}^ \ast  } \left( {\theta ^
\prime ,\phi ^ \prime  } \right)Y_{lm} \left( {\theta ,\phi }
\right) . \label{pato3}
\end{equation}
By means of (\ref{pato3}) and using usual properties for the
spherical harmonics, the term (\ref{ener2}) reduces to the Coulomb
energy after subtracting the self-energy term. On the other hand,
it should be noted that in order to evaluate Eqs. (\ref{ener3})
and (\ref{ener4}) we need to know the magnetic field ${\bf B}({\bf
x})$. However, since we are dealing with an external constant
field $\theta$, we restrict ourselves to constant magnetic fields,
${\bf B}({\bf x})={\bf B}({\bf 0})$. Then, with this assumption
and using (\ref{pato3}), the interquark potential at lowest order
in $\theta$ becomes
\begin{equation}
V =  - \frac{{e^2 }}{{4\pi }}\frac{1}{L}\left[ {1 + 3\left(
{\mbox{\boldmath$\theta$}  \cdot {\bf B}\left( {\bf 0} \right)}
\right) - 6\left( {\mbox{\boldmath$\theta$} \cdot {\bf \hat{r}}}
\right)\left( {{\bf B}\left( {\bf 0} \right) \cdot {\bf \hat{r}}}
\right)} \right]  , \label{ncpotential}
\end{equation}
where $L\equiv\mid {\bf y} \mid$ and ${\bf \hat{r}}_i  =
\frac{{{\bf y}_i }}{{|{\bf y}|}}$.

Accordingly, to lowest order in $\theta$, the nature of the static
potential remains unchanged. Nevertheless, the introduction of the
noncommutative parameter induces a renormalization of the charge,
which is absent in the corresponding ordinary spacetime. In such a
case, Eq. (\ref{ncpotential}) may be rewritten as
\begin{equation}
V =  - \frac{{e_{ren}^2 }}{{4\pi }}\frac{1}{L}. \label{newpot}
\end{equation}

To conclude, the expressions for the corrections to the static
energy obtained from both the Born-Infeld and noncommutative
electrodynamics (to order $e^2$) are quite different. This means
that the two theories are not equivalent. Born-Infeld
electrodynamics has a rich structure which is reflected in a
long-range correction to the Coulomb potential, which is not
present in its noncommutative counterpart. On the other hand, the
present investigation reveals the general applicability of our
methodology. It seems a challenging work to extend to higher
orders the above analysis. We expect to report on progress along
these lines soon.

\section{ACKNOWLEDGMENTS}

Work supported in part by Fondecyt (Chile) grant 1030355.

\end{document}